\documentclass[12pt]{article} 
\usepackage{epsf}

\usepackage{hyperref} 

\DeclareFixedFont{\xiiss}{OT1}{cmss}{m}{n}{12}
\DeclareFixedFont{\ixss}{OT1}{cmss}{m}{n}{9}
\DeclareFixedFont{\cmrnine}{OT1}{cmr}{m}{n}{9}
 
\newcommand{\CC}{\hbox{\xiiss C\kern-.4emI}}
\newcommand{\RR}{\hbox{\xiiss R\kern-.45emI}}
\newcommand{\ZZ}{\hbox{\xiiss Z\kern-.4emZ}}
\newcommand{\CCs}{\hbox{\ixss C\kern-.4emI}}
\newcommand{\ZZs}{\hbox{\ixss Z\kern-.4emZ}}

\newcommand{\beq}{\begin{equation}}
\newcommand{\beql}[1]{\begin{equation}\label{eq:#1}}
\newcommand{\eeq}{\end{equation}}
\newcommand{\be}{\begin{equation}}
\newcommand{\ee}{\end{equation}}
\newcommand{\beqn}{\begin{eqnarray}}
\newcommand{\eeqn}{\end{eqnarray}}
\newcommand{\bea}{\begin{eqnarray}}
\newcommand{\eea}{\end{eqnarray}}

\newcommand{\myfig}[3]{
 \begin{figure}[ht]
 	\begin{center}
		\leavevmode
 		\epsfxsize=#2in
 		\epsfbox{#1}
 	\end{center}
 	\caption{#3}
 	\label{fig:#1}
 \end{figure}
}

\begin{document}
        \begin{titlepage}
        \title{
                \begin{flushright}
                \begin{small}
                ILL-(TH)-98-07\\
                hep-th/9812142\\
                \end{small}
                \end{flushright}
               \vspace{1cm}
			String Junctions and \\
			Bound States of Intersecting Branes
		}

\author{ 	David Berenstein\thanks{email:{\tt berenste@hepux0.hep.uiuc.edu}}
			\\
			and \\
			Robert G. Leigh\thanks{email:{\tt rgleigh@uiuc.edu }}\\ 
        	\\
                {\small\it Department of Physics}\\
                {\small\it University of Illinois at Urbana-Champaign}\\
                {\small\it Urbana, IL 61801}\\
		}

        \maketitle

        \begin{abstract}
We study four-dimensional black hole configurations which result
from wrapping M5-branes on a Calabi-Yau manifold, as well as U-dual
realizations. Our aim is to understand the microscopic degrees of
freedom responsible for the existence of bound states of multiple
branes. The details depend on the chosen U-frame; in some cases,
they are massless string junctions. We also identify a perturbative
description in which these states correspond to twisted strings of
intersecting D3-branes at an orbifold singularity. In each case, these
are the preponderant states of the spacetime infrared conformal field
theory and account for the entropy of the blackhole.
        \end{abstract}

        \end{titlepage}


\section{Introduction}

It is by now well-known that systems of intersecting branes correspond
to blackholes, and the entropy of such a system may be accounted for by
enumerating string states \cite{Strominger:1996sh}. At least when
sufficient supersymmetry is preserved, the configuration of branes is a
bound state at threshold. In many cases, these bound states signal the
existence of degrees of freedom localized on the intersection manifold.
It will be the aim of this note to understand in more detail the nature
of these new states.

We are interested here in an intuitive problem: what is the detailed
mechanism for binding together a collection of many (more than two)
branes, and in particular, what are the relevant microscopic degrees of
freedom? For a bound state of a pair of branes, we can certainly expect
that ordinary strings  stretching between them are responsible for the
binding. However, in intersections of more than two branes, binding by
ordinary strings cannot account for the entropy of the configuration, as
we will discuss in some detail below.

The system that we will have in mind throughout this paper is the
four-dimensional blackhole obtained from an M5-brane wrapped on a
divisor of a Calabi-Yau threefold. However, it will be useful to
consider directly a collection of three types of M5-branes wrapped on
orthogonal cycles of a $T^6$. In much of the paper, we will discuss
directly the case of $T^6$, although we explore Calabi-Yau's in the
final section. In the case of $T^6$, we may take the M5-branes to be
arranged as follows:
\newcommand{\bin}{$\bullet$}\newcommand{\bout}{--}
\begin{center}
\begin{tabular}{c|ccccccccccc} 
	Brane & 0& 10 & 1& 2 &3 &4 &5 &6 & 7 & 8 & 9\\ \hline
	$M5_1$ & \bin & \bin & \bin & \bin & \bin & \bout & \bout & \bin & \bout & \bout & \bout \\ 
	$M5_2$ & \bin & \bin & \bin & \bout & \bout & \bin & \bin & \bin & \bout & \bout & \bout \\ 
	$M5_3$ & \bin & \bout & \bout & \bin & \bin & \bin & \bin & \bin & \bout & \bout & \bout \\
	$P_L$ & \bout & \bout & \bout & \bout & \bout & \bout & \bout & \bin & \bout & \bout & \bout 
\end{tabular}
\end{center}
The four-dimensional blackhole has an $E_{7,7}$ U-duality group; a
useful diagonal basis identifies four charges as the number of M5-branes
of each of three types plus momentum along the eleventh direction. The
entropy of this blackhole is given, at least to leading order, by the
product of these charges, and may be thought of as counting all of the
excitations of the blackhole.

Let us briefly review what is known about this system. There are several
points of view. In the limit where the compact manifold is small, one
attains an effective description in terms of a $1+1$-dimensional field
theory on the intersection manifold of the M5-branes. This theory is a
superconformal field theory in the infrared, with $(0,4)$ supersymmetry.
First, there is an important analysis of Ref. \cite{msw} (see also Ref.
\cite{jrg}) which computes
the central charge of this theory in terms of the cohomology of the
complex divisor upon which the M5-branes are wrapped. Thus the entropy
is computed, in leading order, by the triple-self-intersection number of
the divisor. This number can be thought of as the number of free fields
required to describe the entropy  of the system. String states
stretching between two types of branes would only account for double
intersections, and thus fall short. To our knowledge, a concrete
proposal for the target space of a $\sigma$-model has not been given,
although perhaps intuitively one expects that this then is related to
the complexified moduli space of the divisor. Whatever this spacetime
CFT is, it is known that on $T^6$ it must have a moduli space of
deformations given by $F_{4(4)}(\ZZ)\backslash F_{4(4)}\Big/ Sp(2)\times
Sp(6)$.\cite{kll}

The low energy physics of the bound states may be understood in terms
of deformation theory. Locally, we can discuss the triple intersection
in $\CC^3$, coordinatized by $z^1,z^2,z^3$. An equation for the divisor
is of the form 
\beq
P_{N_1,N_2,N_3}(z^1,z^2,z^3) =0 =
P_{N_1}(z^1) P_{N_2}(z^2) P_{N_3}(z^3)
\eeq
where $N_i$ are the degrees of each polynomial. The zeroes of this 
polynomial correspond to the position of each M5-brane. The holomorphic
deformations of the divisor are of the form
\beq
P_{N_1,N_2,N_3}(z^1,z^2,z^3) + Q_{N_1-1,N_2-1,N_3-1}(z^1,z^2,z^3) = 0
\eeq
The degrees of the polynomial $Q$ have been chosen such that this
deformation does not alter the asymptotic form. The deformations are
localized at triple intersections. To see this, fix $z^{1,2}$ very large
away from the zeroes of the polynomial; it is then clear that the third
variable will be very small. That is, the deformations can only be large
when $z^{1,2}$ are close to the zero of their respective polynomial; this
may be verified explicitly.

We can choose to write the deformations in the following form:
\beq 
Q_{N_1,N_2,N_3} = \sum_{i,j,k} a_{ijk}
\frac{P_{N_1}(z^1) P_{N_2}(z^2) P_{N_3}(z^3)}
{(z^1-r^1_i)(z^2-r^2_j)(z^3-r^3_k)} 
\eeq
The $a_{ijk}$ are the localized deformations, and appear as fields in
the low energy description. The number of degrees of freedom then is
simply counted as the number of triple intersections; because of
supersymmetry, these must come in supermultiplets, with $c=6$. When we
compactify, care must be taken with boundary conditions, and so not all
of these deformations are allowed. One expects, however, that these
effects are subleading compared to the number of triple intersections.
We will see evidence of this below.

Furthermore, the near-horizon limit of this blackhole displays geometry
$AdS_3\times S^3/\ZZ_N\times M_4$; the supergravity spectrum on $AdS_3$
has been computed\cite{finn}, and recently, the quantization of
strings\cite{kll} in this background has been considered. In this paper,
we are not directly interested in such SCFT descriptions. Instead, we
would like to elucidate the microscopic stringy physics responsible for
the existence of the boundstate. The physics that we are interested in
will appear quite different from the point of view of different U-dual
frames. We discuss several different U-frames here; perhaps the most
intuitively appealing picture is within a Type IIB frame, where the
binding of three branes is related to the existence of massless string
junctions localized at the triple intersection. The identification of
these non-perturbative states is hampered by the absence of BPS states
in this background, although we give strong arguments for the existence
of the boundstates. Another Type IIB frame involves intersecting
D3-branes localized at an orbifold singularity; the bound states are
understood in terms of twisted strings. The latter frame leads to a
perturbative UV gauge theory description of this system.


\section{String Junctions}

We begin with a short review of the essential properties of string junctions.
In Type IIB string theory, 1-branes are classified by a pair of integers
$(p,q)$. In this notation, the fundamental string is a $(1,0)$-brane, and
the $D1$-brane a $(0,1)$-brane. It is known that, subject to some conditions,
there is a BPS state consisting of three such branes meeting at a
junction. Since $p$ and $q$ are the charges with respect to the 2-forms
$B_{NS}$ and $B_R$, they must be conserved at the vertex:
\beq
\sum_i p_i=\sum_i q_i =0.
\eeq
In addition, there is a condition on the tensions of the branes, and
this condition depends on the string coupling
\cite{Callan:1998sf,Dasgupta:1997pu}.

Now note that there is a U-duality frame in which the 3 M5-branes become
an NS5-brane, a D5-brane and a D3-brane in Type IIB string theory. This
is attained (refering to the table in Section I) by compactifying the
10-direction, then performing T-duality along, say, the 2-direction.
These three branes intersect along a string as did the M5-branes. The
low energy theory then is expected to be a $1+1$-dimensional CFT with
$(0,4)$ supersymmetry.
\begin{center}
\begin{tabular}{c|ccccccccccc} 
	Brane & 0& 1& 2 &3 &4 &5 &6 & 7 & 8 & 9\\ \hline
	$D3$ & \bin & \bin & \bout & \bin & \bout & \bout & \bin & \bout & \bout & \bout \\ 
	$D5$ & \bin & \bin & \bin & \bout & \bin & \bin & \bin & \bout & \bout & \bout \\ 
	$NS5$ & \bin & \bout & \bin & \bin & \bin & \bin & \bin & \bout & \bout & \bout \\
	$P_L$ & \bout & \bout & \bout & \bout & \bout & \bout & \bin & \bout & \bout & \bout 
\end{tabular}
\end{center}

It is well known that fundamental strings may end on D-branes, and by
S-duality, the D1-brane may end on the NS5-brane. Since the D3-brane is
S-invariant, any $(p,q)$-1-brane may end on it. Thus, at least from the
point of view of charge conservation, the state shown in Figure 1
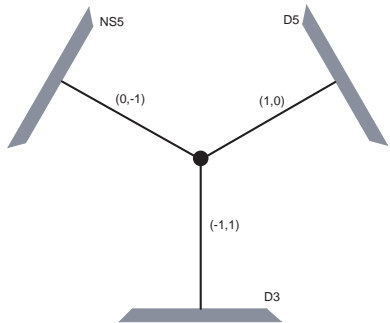
exists. \myfig{junctionbind.EPSF}{2}{Cartoon of junction between
branes.} Furthermore, the string junction is massless when the three
branes intersect; the junction may be made massive by moving the branes
away from each other in the 789-directions.

Now, each of the ends of the string junction may terminate on any of the
$N$ branes of the appropriate type. Thus, we see that there are of order
$N_1N_2N_3$ states present here. Furthermore, since the junction must
organize itself into a representation of the $(0,4)$ supersymmetry,
there are $4N_1N_2N_3$ bosonic states and their superpartners. String
junctions then account for the entropy of this configuration. Note that
in this frame, open string states stretching between branes are not this
numerous. Thus, at least to leading order, the entropy is accounted for
by non-perturbative states.

There are several potential problems with this picture however, and we
now turn to a discussion of the relevant issues. We have claimed above
that the string junctions are massless when the branes intersect.
Although this is clearly true geometrically at the classical level, it
is not true that the mass of a massive state is protected. To understand
the relevant issues, we should consider the details of $(0,4)$
supersymmetry algebra in two dimensions.\cite{Witten:ADHM} The algebra
takes the form
\beq
\{ Q, Q\} = P_R
\eeq
In particular,  there are no central charges as that requires both left
and right moving supersymmetries. The BPS bound is thus simply $P_R\geq
0$; the only states saturating the bound are {\it massless} and may have
$P_L\neq 0$. This implies that in any ultraviolet description, only the
massless states with $P_R=0$ will necessarily survive down to the
infrared conformal theory and contribute to the entropy of the
configuration we are studying. For this massless state to be present
then, we must argue that the classical moduli space is unmodified
quantum mechanically, at least at the origin. Indeed, we do not expect
such modifications because of the $ (0,4) $ supersymmetry. This is
actually more restrictive than
$ (2,2) $; for example, the metric of the target space manifold must
be hyperk\"ahler. Further evidence will be presented below.

If we identify the states localized at the intersection to be of a
non-perturbative origin (at least in this frame), then we must become
comfortable with the idea that the conformal field theory of ordinary
string states is somehow insufficient. Indeed, we can think of this
situation as akin to a conifold singularity--at the origin, there is a
new branch of the moduli space, parameterized by vev's of the fields
corresponding to string junctions. This is not obviously inconsistent, as
near the NS5-five branes the string theory is strongly coupled which
invalidates perturbation theory.

In the next section, we consider a different U-frame, in which these
states appear in the perturbative spectrum.


\section{The Orbifold Frame}

In this section, we discuss another U-frame which is perturbative, and
the localized states at the intersection are twisted strings. To attain
this, we may begin with the configuration of the last section, and 
perform a T-duality along $X^{1,2}$:
\begin{center}
\begin{tabular}{c|ccccccccccc} 
	Brane & 0& 1& 2 &3 &4 &5 &6 & 7 & 8 & 9\\ \hline
	$D3_1$ & \bin & \bout & \bin & \bin & \bout & \bout & \bin & \bout & \bout & \bout \\ 
	$D3_2$ & \bin & \bout & \bout & \bout & \bin & \bin & \bin & \bout & \bout & \bout \\ 
	$KK5 $ & \bin & $\times$ & \bin & \bin & \bin & \bin & \bin & \bout & \bout & \bout 
\end{tabular}
\end{center}
The interpretation of this configuration is that of a pair of D3-branes
intersecting along a line ($X^6$), at a $\ZZ_{N_3}$ orbifold
singularity.\footnote{In the table, the symbol $\times$ refers to the 
Taub-NUT direction. We take $X^{1,7,8,9}$ to be noncompact. This
ensures that the singularity is isolated.} Here, $N_3$ is the number of
$NS5$-branes in the original picture, and there are $N_1$ ($N_2$)
D3-branes of each type. Note that in this frame, there is no manifest
triality between $N_1,N_2$ and $N_3$. This occurs simply because of
taking a definite U-duality frame; triality will be recovered in
U-invariant quantities, such as the entropy.

This is an interesting configuration in its own right. There has been
several appearances of D3-branes at orbifold singularities in the
literature, giving rise to interesting $3+1$-dimensional gauge theories.
In the present configuration, we find a gauge theory description of the
$1+1$-dimensional intersection. This theory is an ultraviolet
description where gravity has been decoupled, which will flow to the
relevant conformal field theory in the infrared. In this theory, we will
be able to identify the states that are localized at the intersection,
and which contribute the predominant amount of entropy. Since the
configuration is perturbative, the analysis is reliable. Furthermore, we
will be able to map these states to the string junctions of the previous
section.

The spectrum of this gauge theory may be obtained via a straightforward
application of familiar techniques. Note first that if we concentrate on
the states of a single D3-brane but dimensionally reduce along a two
torus, we expect to see multiplets of $(4,4)$ supersymmetry. The
supersymmetry preserved by each of the two D3-branes is incompatible,
and at the end we are only left with $(0,4)$ supersymmetry; the string
states connecting $D3_1$ to $D3_2$ do not form full $(4,4)$-multiplets.
In fact, we will find that the orbifolding acts as to shift the gauge
quantum numbers of fermions with respect to those of bosons.

To construct the spectrum, account for the orbifolding by $N_3$ images
of the collections of $N_1$ ($N_2$) D3-branes. String states that
stretch between D3-branes of the same type, as mentioned, give
multiplets of $(4,4)$ supersymmetry--the fermions and bosons are in the
same gauge multiplets. Those multiplets which correspond to string
states between branes at the same image, turn out to be hypermultiplets,
whereas those stretching horizontally (see Fig. \ref{fig:quiver.EPSF})
are vector multiplets, (this nomenclature comes from looking at the four
dimensional theory on the intersection of two D5-branes, where the
vector directions are along the intersection manifold, and the
hypermultiplet directions are orthogonal). The resulting gauge group is
then\footnote{We consider the low energy ultraviolet theory, and so do
not concern ourselves with the possible decoupling of $U(1)$'s.}
\beq
\prod_{k=1}^{N_3} \left[ U(N_1)\times U(N_2)\right].
\eeq

The string states that stretch between D3-branes of different type
however are acted upon non-trivially by the orbifold. It should be noted
that the $\ZZ_{N_3}$ acts chirally on the $SU(2)\times SU(2)$ R-symmetry
on either of the D3-branes. There are several reasons for this choice.
First, this particular orbifold action is important for preserving
$(0,4)$ supersymmetry and the resulting hyperk\"{a}hler structure. More
importantly, the corresponding 4-dimensional blackhole is, as in Ref.
\cite{kll}, related to a configuration of NS5-branes and KK monopoles,
for which the near-horizon geometry is given by $AdS_3\times
S^3/\ZZ_{N_3}$. In the near-horizon region, the orbifold of the sphere
indeed acts chirally. This is no coincidence; in fact both the
near-horizon geometry, as well as this gauge theory description, share
the same geometrical features. The gauge theory, then, is an ultraviolet
description of the spacetime conformal field theory which controls the
physics of the near-horizon region of the blackhole. The detailed form
of this CFT, as mentioned, is not known; however, at the very least, the
gauge theory discussed here should be capable of reproducing some of the
features of the CFT, in particular the chiral ring.\footnote{Note that
in order to check non-chiral operators, we would need to control the
non-perturbative details of the gauge theory in the infrared limit.} We
do not attempt to demonstrate this here.

Given this orbifold action, bosons and right moving fermions form
supermultiplets, and the left moving fermions are singlets under
supersymmetry. The field content is summarized in Fig.
\ref{fig:quiver.EPSF}. The fields are supermultiplets for the vertical
lines, and left-moving fermions for the diagonal lines. The nodes and
edges have a supermultiplet and left-moving fermion singlets, as is
required in order to complete representations of (4,4) supersymmetry.
Note that this portion of the spectrum is an example of ``misaligned
supersymmetry" of Ref. \cite{Dienes}, as bosons and fermions are
degenerate but they are in different representations of the symmetry
groups. Thus much of the structure of a $(4,4)$-supersymmetric theory is
present; only the gauge representations are aware of the breaking to
$(0,4)$.

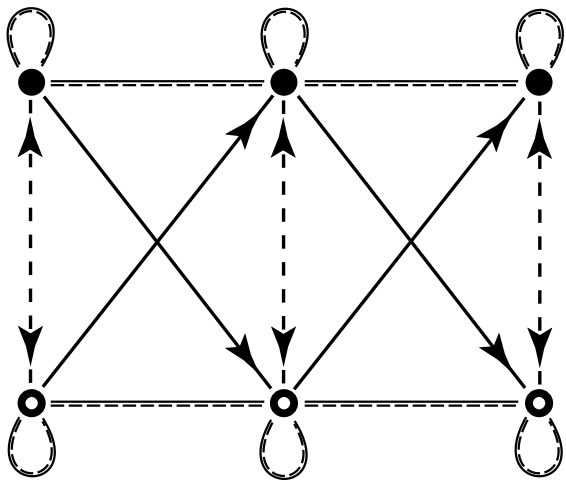
\myfig{quiver.EPSF}{3}{A portion of the quiver diagram. The open
(closed) circles represent images of the $N_1$ ($N_2$) collections of
$D3$-branes. Bosonic string states and superpartners are represented by 
dashed lines, left-moving fermions by solid lines.}

As the configuration is made only out of D3-branes, the value of the
type IIB coupling constant is not fixed at any value, and we can
actually take a weakly coupled limit, so that the field theory analysis
is accurate.

Next, we would like to count (gauge invariant) modes, in order to probe
the entropy of the corresponding blackhole. To facilitate this, we move
on the moduli space to a generic point, where the gauge group is broken
as much as possible. To this end, we move all D3-branes apart in the
2345-directions (but not away from the orbifold singularity). The gauge
group is Abelian, $U(1)^{N_1}\times U(1)^{N_2}$, and massless charged states are
present. While most of the
$(4,4)$ vectors and hypermultiplets have been lifted, the twisted states
survive. These states are localized at the orbifold singularity, and
have multiplicity $N_1N_2N_3$  (since they are in $(N_1,\bar N_2)$
representations, and there are $N_3$ images). For each of these, we have
two complex bosonic modes and two complex fermionic modes (as in Fig.
\ref{fig:quiver.EPSF}). In a sector with fixed $P_L$, these states
dominate the entropy, giving by a standard argument $S= 2\pi
\sqrt{6N_1N_2N_3P_L+\ldots} $. 

It was found in Ref. \cite{kll} that the central charge of the spacetime
conformal theory contains no subleading corrections. However, in the
present construction, it appears that there is a problem. There are
massless fields which are the remnant of the adjoint hypermultiplets.
There is one such supermultiplet remaining per vertex of Fig.
\ref{fig:quiver.EPSF}, and thus one would expect that these fields
contribute to the entropy at order $(N_1+N_2)N_3$. It is possible that
the correct central charge is nevertheless obtained as follows, by
canceling this contribution. We have assumed that all triple
intersections contribute an independent supersymmetric degree of freedom
to the entropy, but this is not really true, as not all of the local
deformations can produce a smooth manifold. This means that some
fraction of the (vertical) fields have a superpotential and therefore do
not contribute to the entropy. It is quite possible that this correction to
the leading term in the entropy above precisely cancels the effect of
the adjoint fields. A similar mechanism is known to occur in the D1-D5
system\cite{jmthesis,handw}--the dimension of the moduli space is smaller
than the number of fields because of D-term constraints (in our
case we have F-terms). It is clear then that the present description is far
from being a free CFT, at least at finite $N$. The gauge theory description
is useful however in the long string limit, where these effects are subleading.
A ueful application would be the computation of the chiral ring.

\subsection{Relation to String Junctions}

Now note that we expect that this discussion of the spectrum is robust--
the entropy is accounted for by twisted string states, as long as the
singularity itself is not modified by quantum corrections. Furthermore,
this description of the states localized at the intersection is T-dual
to the description in the previous section in terms of string junctions,
this is, from one description to the other we do a discrete Fourier
transform. We regard this then as definitive evidence (if duality is to
be believed) for the existence of massless string junctions in that
frame, and hence for their contribution to the entropy.


\section{Other U-frames}

It is of interest to consider other U-frames in the same context. We
confine ourselves to brief discussions of three such frames; in most
cases, an understanding of the localized states is considerably more
difficult.

\subsection{M-theory and 3 M5-branes}

First, we consider the original M-theory configuration, and account for
the entropy there.\cite{igor} This may be understood by beginning with the string
junction; if we lift this to M-theory, we find that the junction becomes
a M2-brane "pants section". Each $(p,q)$-leg has one direction wrapped
along the vector $(p,q)$ in the $X^{2,10}$ torus \cite{Kl}. Thus the
bound state degrees of freedom are these pants sections; at a triple
intersection point of the M5-brane, they have zero area and so should go
massless. A smooth point in moduli space then is attained by turning on
vevs for these low energy fields.

\subsection{Type IIA and the $4440$-system}

By compactifying the M-theory configuration along $X^6$, we obtain a
system of three different types of D4-branes, plus D0-branes from
momentum along $X^6$. This is a system that has been well-studied in the
blackhole context\cite{vjbfl,bll}. The pants section of the preceding
paragraph descends to a similar D2-brane, while the momentum descends to
a constant D0-flux through the D2-brane, $F_2=P_ L dVol$; where the
volume is normalized to unity.

The localized states of this system are counted as follows. At a given
intersection, a pants section is massless,  and  with an arbitrary
D0-flux it's energy is just the D0 brane charge. States with fixed
D0-flux $P_L$ are then obtained by partitioning that flux over $n$ pants
sections (where $1\leq n \geq P_L$, in the normalization where $P_L$ is
an integer.) Thus we find factorial growth of states exactly like the
free field theory calculation, with\footnote{This may be obtained by
taking the partition function $Z\simeq \left(
\eta(q)\theta(q)\right)^{4N_1N_2N_3}$, for fixed $\langle
E\rangle=N_0$.} $S\simeq 2\pi\sqrt{6N_0N_1N_2N_3}$. Notice that the $D2$
branes are in a sense auxiliary to the construction as the total $D2$
brane charge is zero.


\subsection{Type IIB and the $3333$-system}

If we T-dualize the Type IIA configuration along three directions, such
as $X^{1,3,5}$, we find four different types of D3-branes, which now
intersect at a point:
\begin{center}
\begin{tabular}{c|ccccccccccc} 
	Brane & 0& 10 & 1& 2 &3 &4 &5 &7 & 8 & 9\\ \hline
	$D3_1$ & \bin & \bin & \bout & \bin & \bout & \bout & \bin & \bout & \bout & \bout \\ 
	$D3_2$ & \bin & \bin & \bout & \bout & \bin & \bin & \bout & \bout & \bout & \bout \\ 
	$D3_3$ & \bin & \bout & \bin & \bin & \bout & \bin & \bout & \bout & \bout & \bout \\ 
	$D3_4$ & \bin & \bout & \bin & \bout & \bin & \bout & \bin & \bout & \bout & \bout 
\end{tabular}
\end{center}
The fourth D3-brane comes from the D0-branes of the Type IIA frame. Note
that each pair of these D3-branes intersects along a line, but the
four branes intersect at most at a point. Thus, the
low energy description would again be a $1+0$ quantum mechanical
system. 

In this U-frame, a description of the bound states appears to be very 
complicated.
To see this consider the transformation of the string junction under
the T-duality mentioned above. Depending upon which three directions
that we T-dualize along, we get a pants section of D5-branes, or
D3-branes, or a mixture of the two. It would seem counterintuitive
to attempt an explanation of bound states of D3-branes in terms
of D5-branes! However, we note that there are global conditions that
must be satisfied to maintain charge conservation. These conditions
(vanishing of total brane charge) imply
that the description of the bound state in terms of, say, D5-branes
is unstable. Perhaps there is a description of these bound states in
terms of some remnant, along the lines of Refs. \cite{sen,ewrem}.


\section{M-branes on Calabi-Yau Threefolds}

It is also of interest to discuss the case of M5-branes on Calabi-Yau
3-folds more directly. To begin, consider the case of $K3\times T^2$,
with M5-branes wrapped on different complex 2-cycles. In particular,
there are M5-branes wrapped on the whole $K3$ manifold; in a Type IIB
description, these branes give rise to an $A_{N_3}$ singularity times a
$K3$ surface. Other M5-branes that wrap the $T^2$ as well as a 2-cycle
of the $K3$ correspond to D3-branes. Again, we can go to a weakly
coupled type IIB picture and repeat the steps to get the open string
quiver diagram corresponding to the configuration. The twisted open
strings are again the relevant degrees of freedom.

This construction can be immediately generalized to an elliptically
fibered Calabi-Yau manifold $M$ with a section. Clearly, we should
distinguish M5-branes which wrap a cycle on the base plus the elliptic
fibre from those which wrap the base completely. The latter appear as KK
monopoles while the former become D3-branes wrapped on 2-cycles of the
base, once we turn to the IIB F-theory configuration.

Therefore, we expect a {\it local} description as D3-branes wrapping
cycles of the base at an orbifold singularity. The description in terms
of twisted open strings should still be good locally on the D3 branes,
yet the choice of which string is light changes as we move  around the
D3 brane, and they certainly become massless at the intersection points
of these.


\section{Concluding remarks}

In this note, we have considered configurations of branes which form
bound states at threshold. The entropy of these objects may be
understood from the counting of (not necessarily perturbative)  states
which becomes massless when the different constituents of the black hole
are brought together. The identification of these modes as string
junctions is particularly appealing, as all of the degrees of freedom
can be seen geometrically, but are never perturbative in this U-frame.

We have also found a perturbative picture in which the microscopic
states are twisted string states on the intersection of D3-branes at an
orbifold singularity. The ultraviolet theory then is a gauge theory. We
have been unable, by deforming the moduli space, to find a description
of the spacetime infrared conformal field theory in terms of free fields
however, either on the torus, or for those Calabi-Yau manifolds for
which the construction makes sense. It is expected however that this
construction is capable of reproducing the chiral ring.

\noindent {\bf Acknowledgments:} We wish to thank F. Larsen for
discussions. Work supported in part by the United States Department of
Energy grant DE-FG02-91ER40677 and an Outstanding Junior Investigator
Award.

\providecommand{\href}[2]{#2}\begingroup\raggedright\endgroup

\end{document}